\documentclass[twocolumn,letterpaper,aps,prc,showpacs,floatfix]{revtex4}
\usepackage{amsmath,bm}
\usepackage{graphicx}
\begin{document}
\vspace{1.8in}
\title{Hadron production from quark coalescence and jet fragmentation  \\ 
in intermediate energy collisions at RHIC}
\author{V. Greco}
\affiliation{Cyclotron Institute and Physics Department, Texas A\&M 
University, College Station, Texas 77843-3366, USA}
\author{C. M. Ko}
\affiliation{Cyclotron Institute and Physics Department, Texas A\&M 
University, College Station, Texas 77843-3366, USA}
\author{I. Vitev}
\affiliation{Theory Division and Physics Division, 
Los Alamos National Laboratory, Mail Stop H846, 
Los Alamos, New Mexico 87545, USA}
\date{\today}

\begin{abstract}
Transverse momentum spectra of pions, protons and antiprotons in Au+Au
collisions at intermediate RHIC energy of $\sqrt{s_{NN}}=62$~GeV are
studied in a model that includes both quark coalescence from the
dense partonic matter and fragmentation of the quenched perturbative 
minijet partons. The resulting baryon to meson ratio at
intermediate transverse momenta is predicted to be larger than 
that seen in experiments at higher center of mass energies.
\end{abstract}

\maketitle

\section{introduction}
Recently, there is a renewed interest in using the quark coalescence
or recombination model to study hadron production in ultra-relativistic 
heavy ion 
reactions~\cite{voloshin,hwa,fries,greco,greco2,moln,lin03,greco-res,greco-c}. 
These studies were largely triggered by two surprising observations   
in measured hadron spectra from Au+Au collision at
$\sqrt{s_{NN}}=200$~GeV at the Relativistic Heavy Ion Collider 
(RHIC)~\cite{voloshin,ppi,adlv2,adams}.  The first was the nearly similar 
yield of protons and pions at  intermediate transverse momentum 
(2~GeV $ < p_T < $ 5~GeV), which is at variance with the expectation
from the fragmentation of minijets produced from initial hard
processes. The second was the so-called ``quark number scaling'' of
hadron elliptic flow $v_{2h}(p_T)$, i.e., the transverse momentum
dependence becomes universal if both $v_{2h}$ and transverse momentum 
$p_T$ are divided by the number of constituent quarks in the hadron. Both 
phenomena have a simple explanation if hadronization of the thermally 
equilibrated and collectively flowing partonic matter formed in these 
collisions proceeds through the coalescence of quarks of constituent 
masses~\cite{greco,greco2,fries,moln}.  

The description of hadronization of the quark-gluon plasma formed in 
heavy-ion collision at relativistic energies by quark coalescence was
first introduced in models like ALCOR~\cite{alcor} and
MICOR~\cite{micro}. Emphases of these earlier studies were on particle
yields and their ratios. Recent studies have been more concerned with 
observables that are related to collective dynamics and production of 
hadrons with relatively large transverse momenta.  Furthermore,
effects of minijet partons from initial hard processes have also been 
included through their independent fragmentation as well as
coalescence with soft partons in the quark-gluon 
plasma~\cite{greco,greco2}. The interplay between quark 
coalescence and minijet fragmentation has been found to be essential 
for understanding measured inclusive hadron spectra at moderate and 
high transverse momenta. If confirmed by more exclusive data from 
experiments and by further theoretical studies, hadronization via 
quark coalescence then represents a new probe of the quark-gluon
plasma formed in relativistic heavy ion collisions. It is thus of
great interest to study the onset of the coalescence mechanism as 
the collision energy  is varied. Also, comparison of results from the
coalescence model for heavy ion collisions at different energies with 
upcoming experimental data will allow for a better understanding of
its predictive power.
In the present paper, such a study is carried out by using the same 
coalescence model as applied previously to higher energy collisions
to make predictions for the proton, antiproton, and pion spectra at 
a lower collision energy of $\sqrt{s_{NN}}=62$~GeV. 

This paper is organized as follows. In Sect.~\ref{coalescence} we 
briefly review the coalescence model used in Ref.~\cite{greco}. How the
parameters in the model are determined for Au+Au collisions at 
$\sqrt{s_{NN}}=62$~GeV is described in Sect.~\ref{parameters}. We
then discuss in Sect.~\ref{pQCD} the fragmentation contribution from 
a pQCD approach that takes into account both the Cronin effect and the 
radiative in-medium energy loss~\cite{vitev-62}. In
Sect.~\ref{results}, results on pion, proton and antiproton spectra 
and their ratios are presented. Finally, a summary is given in 
Sect.~\ref{summary}. 
  
\section{the coalescence model}\label{coalescence}

The coalescence model describes the dynamical process of converting quarks
and antiquarks into hadron bound states in the presence of  partonic matter.
In general, the process of coalescence would involve the emission of a third
particle or, more generally, off-shell effects to guarantee energy
conservation and color neutrality. However, under the assumption of a 
fast process in which the binding interaction is turned on suddenly, 
the probability of forming the bound state is simply given by the overlap 
of the quark and antiquark distribution functions with the Wigner function 
of the formed hadron. In this case, the transverse momentum spectrum 
of a hadron consisting of $n$ (anti-)quarks can be written as  
\begin{eqnarray}
\frac{dN_{H}}{d^2P_T}&=&g_{H} \int \prod_{i=1}^{n}
\frac{d^{3}\mathbf{p}_{i}}
{(2\pi)^{3}E_{i}}{p_{i}\cdot d\sigma _{i}}f_{q}(x_{i},p_{i})\,\nonumber\\
&\times&f_{H}(x_{1}..x_{n};p_{1}..p_{n})\delta^{(2)}
\left(P_T - \sum_{i=1}^n p_{T,i}\right). 
\label{coal}
\end{eqnarray}
In the above, $d\sigma$ denotes an element of a space-like hypersurface,
and $g_H$ is the statistical factor for forming a colorless hadron
from colored quarks and antiquarks of spin 1/2. For hadrons considered 
here, i.e., $\pi$, $\rho$, $K^*$, $p$, and $\Delta$, the statistical
factors are, respectively, $g_\pi=g_\omega=1/36$, $g_\rho=g_{K^*}=1/12$, 
$g_p=1/108$, and $g_\Delta=1/54$. The $n$-quark phase space
distribution function has been approximated in Eq.~(\ref{coal}) by the 
product of the single quark distribution function $f_q(x,p)$ that is 
normalized to the quark number in the partonic matter.

The Wigner function $f_H(x_i;p_i)$ in Eq.~(\ref{coal}) describes
the spatial and momentum distributions of quarks in a hadron. 
For mesons, we have taken the Wigner functions to be a
sphere of radius $\Delta_x$ in coordinate space and 
$\Delta_p=\Delta_x^{-1}$ in momentum space, i.e., 
\begin{eqnarray}
&&f_M(x_1,x_2;p_1,p_2)=\frac{9\pi}{2}
\Theta\left(\Delta_x^2-(x_1-x_2)^2\right)\nonumber\\
&&\times\Theta\left(\Delta_p^2-(p_1-p_2)^2+(m_1-m_2)^2\right).
\label{wigner}
\end{eqnarray}
This represents the simplest ansatz for the Wigner function, i.e, 
it gives a probability of one if the relative covariant distances
between two particles in space and momentum are smaller than the
radius parameters. The Wigner functions for baryons can be 
similarly expressed in terms of appropriate relative coordinates 
(see Ref.~\cite{greco2} for detail).  Although hadron wave functions
are generally not known in detail, the particular shape of
the wave function used in our model does not affect significantly the
result on hadron transverse momentum spectra, unless very extreme
choices are taken. 
As in our previous work~\cite{greco}, we take $\Delta_p= 0.24$~GeV for 
mesons and $\Delta_p= 0.36$~GeV for baryons in the present study. 

\section{Fireball parameters}\label{parameters}

One of the main ingredients in our model is the parton distribution
functions at hadronization. They can in principle be determined from a 
dynamic approach such as the transport 
model~\cite{ampt,moln-MPC,moln04,muller,greiner}.
To focus on the hadronization process itself, we assume instead that at
hadronization the partonic matter consists of both a thermally and
chemically equilibrated quark-gluon plasma and quenched minijet 
partons from initial hard processes.  
Specifically, the single quark distribution function is taken to
consist of both a thermal and a minijet component, separated by the 
transverse momentum $p_0 \sim 2$~GeV. In this section we discuss how
the thermal component is determined from the measured total transverse
energy as in our previous work at $\sqrt{s_{NN}}=200$~GeV~\cite{greco2},
while the minijet one will be discussed in the next section together
with the minijet fragmentation process.

Under the assumption that the quark-gluon plasma is thermally
equilibrated and collectively flowing, the determination of its
parton distributions requires the knowledge of the radial flow and its
total transverse energy. A blast wave analysis of the experimental
data shows that the collective radial flow differs very little between 
the highest SPS and RHIC energies. We can therefore assume the same 
radial flow for Au+Au collisions at $\sqrt{s_{NN}}=62$~GeV as at
higher RHIC energies, i.e., $\beta_{max}=0.5$ or a slightly smaller 
one of about 0.45 and the usual linear radial profile 
$\beta=\beta_{max}\, r/R$, where $R$ is the transverse size of the 
fireball. We have checked that this uncertainty does not have a
significant effect on the transverse momentum dependence of the
$p/\pi$ ratio, which is the main observable we want to study in this paper.

As to the total transverse energy, we are not aware at the moment of 
any measurement for Au+Au collisions at $\sqrt{s_{NN}}=62$~GeV. It has
been, however, noticed that the ratio of the transverse energy to the
total number of charged particles in relativistic heavy ion collisions
is nearly constant in the energy range 20~GeV $< \sqrt{s_{NN}}<$ 200
GeV, and available experimental data \cite{phenix-et} can be 
parameterized by 
\begin{equation}
\frac{dE_T}{dN_{ch}}=0.88+1.44\cdot 10^{-3}(\sqrt{s_{NN}}-130)\, {\rm~GeV}.
\label{detran}
\end{equation}
Also, the charged particle multiplicity normalized to the number of 
participants has a smooth logarithmic behavior that can be parameterized as
\begin{equation}
\frac{dN_{ch}}{d\eta (N_{part}/2)}=0.37 + 0.62\, \ln \sqrt{s_{NN}}. 
\label{dncharged}
\end{equation}
Using the Glauber model to estimate the number of participants, 
we have $N_{part}=330$ for the $0-10\%$ centrality bin.  From
Eq.~(\ref{dncharged}), the charged particle multiplicity is then
$dN_{ch}/d\eta = 480$, corresponding to a multiplicity per unit of 
rapidity of $dN_{ch}/dy= 1.2\, {dN_{ch}}/{d\eta}=580$,
with $d\eta/dy=1.2$ from the relation between rapidity and
pseudorapidity. Eq.~(\ref{detran}) thus leads to a transverse energy 
$dE_T/dy \simeq 460$~GeV for central Au+Au collisions at
$\sqrt{s_{NN}}=62$~GeV.

What remains to be determined is the relative abundance of quarks and 
antiquarks, 
which is essentially
given by the baryon chemical potential. This can in principle also be 
extracted from experimental data on the $\bar p/p$ ratio via 
$\mu_q = - T/6\,\ln(N_p/N_{\bar p})$. For collisions at 
$\sqrt{s_{NN}}=200$~GeV, we already had an experimental value of
$\bar p/p\simeq$ 0.7, which corresponds to a quark chemical potential 
$\mu_q = 11$ MeV~\cite{greco2}. There is no measurement of the $\bar
p/p$ ratio yet at $\sqrt{s_{NN}}=62$~GeV, but it can be estimated
from a parameterization of the chemical potential in the energy
range of 10~GeV $< \sqrt{s_{NN}}<$ 200~GeV, based on available data
from AGS to RHIC. The resulting quark chemical potential is $\mu_q=23$ MeV,
leading to an antiproton to proton ratio $\bar p/p$= 0.44. Because of 
strangeness conservation, numbers of strange and anti-strange quarks
are equal, i.e., zero strangeness chemical potential. The relative
abundance of strange to light quark is then simply determined by their 
constituent quark masses that are taken to be $m_{u,d}=300$ MeV and
$m_s=475$ MeV. Assuming a critical temperature of $T_c=170$ MeV, 
the partonic matter at hadronization is found to have $N_{u+d}=285$, 
$N_{\bar u+\bar d}=218$, and $N_{s}= N_{\bar s}=79$ quarks and
antiquarks distributed in a volume V=555 $\rm{fm^{3}}$ in order to
have the estimated transverse energy of 460~GeV.  This leads to a
critical energy density of about 0.8 $\rm{GeV/fm^3}$, similar to that
from the lattice QCD calculations~\cite{lattice}.  

\section{Quenched Perturbative QCD partons and hadrons}
\label{pQCD}

Partons with transverse momenta $p_T > p_0$ (few GeV)  are mainly
produced from initial hard collisions between nucleons. 
Their distributions can be determined from the pQCD calculation that 
includes the Cronin effect and radiative energy loss as 
described below.

In ``elementary'' hadron-hadron ($N+N$) collisions, high $p_T$ minijet
parton production can be evaluated using the perturbative QCD factorization 
approach~\cite{collins}. This leads to a differential cross section 
that is expressed as a convolution of the measured parton distribution
functions~\cite{pdfs} $f_{\alpha/N}(x_\alpha,Q_\alpha^2)$ for the
interacting partons ($\alpha = a,b$) with the elementary parton-parton 
cross sections $d\sigma^{(ab \rightarrow cd)}/d\hat{t}$~\cite{owens}:
\begin{eqnarray}
\frac{d\sigma_{jet}}{dy d^2 {\bf p_T}}&=&
K_{NLO}  \sum_{abcd} \int\! dx_a 
dx_b \int d^2{\bf k}_{{\rm T}a} d^2{\bf k}_{{\rm T}b} \nonumber \\
&\times& f({\bf k}_{{\rm T}a})f({\bf k}_{{\rm T}b})
f_{a/p}(x_a,Q^2_a) f_{b/p}(x_b,Q^2_b) \nonumber \\
&\times&\frac{\hat{s}}{\pi}\frac{d\sigma^{(ab\rightarrow cd)}}
{d{\hat t}} \delta(\hat{s}+\hat{u}+\hat{t}). 
\label{hcrossec}
\end{eqnarray}
In the above, $x_a$ and $x_b$ are the initial momentum fractions
carried by the interacting partons. For nucleon-nucleon collisions,  
we use phenomenological smearing  $\left\langle {\bf k}_{\rm T}^2 
\right \rangle = 1.8$~GeV$^2$/c$^2$ via a normalized Gaussian 
distribution function $f({\bf k}_{\rm T})$ to mimic next-to-leading
order and soft gluon resummation effects. The determination of 
$K_{NLO}$ in the partonic cross sections in the absence of initial and 
final state multiple scattering is done as in~\cite{vg-quench}. 

In the presence of cold and hot nuclear matter the multiple elastic, 
inelastic and coherent scattering of the incoming and outgoing 
partons modify the perturbative cross sections Eq.~(\ref{hcrossec}). 
The calculated corrections can be systematically incorporated 
in the pQCD factorization approach~\cite{pqcdms}. For example, 
transverse momentum diffusion in $p+A$ collisions leads to the Cronin 
effect \cite{vg-quench}. In $A+A$ reactions the dominant nuclear effect 
is the energy loss of energetic jets propagating through the  
quark-gluon plasma and the dominant energy loss mechanism is via 
medium-induced gluon bremsstrahlung. To first order in the mean number
of scatterings, the radiative spectrum and the average energy loss, 
calculated in the Gyulassy-Levai-Vitev approach~\cite{gyulassy1} 
for the case of Bjorken expansion~\cite{expand}, reads
\begin{eqnarray}
\label{boostsp} 
{\omega  \frac{dN^g_{ind.}}{d\omega} }
&\approx&  C_R \frac{9\pi \alpha_s^3}{4} \frac{1}{A_\perp} 
\frac{dN^{parton}}{dy} \nonumber \\ 
&\times& \left\{  \begin{array}{ll} 
\displaystyle    \frac{ L  }{ \omega }  + \cdots, \quad  &
\displaystyle \omega \gg {\mu^2 L } / {2 }    \\[2ex]
\displaystyle \frac{ 6 }{\pi \mu^2} 
\, \ln \frac {\mu^2 L }{2 \omega} 
+ \cdots, \quad  &
\displaystyle  \omega \ll {\mu^2 L}/{2 }   
\end{array} \right. \; \\
\frac{ \Delta E }{E}&\approx&  \frac{1}{E} \,  
{C_R} \frac{ 9 \pi \alpha_s^3  }{4} \frac{1}{A_\perp} 
\frac{dN^{parton}}{dy} \, L \nonumber\\
&\times& \left( \ln   \frac{2 E}{\mu^2 L}  + 
\frac{3}{\pi} + \cdots \right),
\label{booste} 
\end{eqnarray}   
where $C_R=4/3 \; (3)$ for quark (gluon) jets, $A_\perp$ is the
transverse size of the interaction region, and $L$ is the distance
traversed by the jets. We note that for the purpose of calculating the 
attenuation of the single inclusive hadronic spectra 1+1D expansion 
approximates well the 3+1D result.  

Eqs.~(\ref{boostsp}) and~(\ref{booste}) illustrate qualitatively
the behavior of the energy loss as a function of the system size, 
density and partonic energy. Quantitative jet quenching analysis
requires numerical evaluation of the the radiative spectra and the
mean energy loss as in Refs.~\cite{vitev-62,vg-quench}. 
In the Poisson approximation of independent gluon emission, the probability 
$P(\epsilon)$ for fractional energy loss $\epsilon=\sum_i\omega_i/E$ 
due to multiple bremsstrahlung can be generated from the radiative
spectrum as in~Ref.\cite{levai1}.  Along the lines of Ref.~\cite{vitev-62},
we choose a suppression factor that corresponds to an initial parton 
density $dN_{parton}/dy=725$. The resulting quenched minijet
distributions are used for the parton distributions at $p_T > p_0$ 
in the coalescence process.  

Since the probability for a minijet parton to coalesce with other
partons to form a hadron is quite small, independent fragmentation
remains the dominant mechanism for hadronization of minijets~\cite{greco2}.
This contribution is obtained by introducing the fragmentation 
function $D_{h/c}(z,Q^2_c)$ for minijet parton $c$ into a hadron of 
flavor $h$, i.e., 
\begin{eqnarray}
\frac{dN_{had}}{dyd^2{\bf p_T}}=\sum_{jet}\int dz_c 
\frac{dN_{jet}({\bf p_T}/z_c)}{dyd^2{\bf p_T}}
\frac{D_{h/c}(z_c,Q_c^2)}{z_c^2},
\end{eqnarray}
where $z_c=p_h/p_c$ is the momentum fraction carried by the observed 
hadron and $Q_c=p_h/z_c$ is the momentum scale for the fragmentation 
process. It is this perturbatively computed hadron spectrum that we
refer as fragmentation component of the hadron spectrum. 

For pions, we use the fragmentation functions of Ref.~\cite{ffs} that are
determined from $e^+e^-$, $ep$, and $p\bar p$ collisions.  For $pp$ 
collisions at $\sqrt{s_{NN}}=62$ GeV, where both the Cronin and
radiative energy loss effects are absent, the predicted pion spectrum 
reproduces very well the experimental data at $p_T\geq$ 3 - 4~GeV but 
overestimates that at $p_T=$ 2 GeV by about a factor of 2. To also
reproduce the pion yield at $p_T \simeq$ 2~GeV, we have thus modified the
minijet parton distribution at low $p_T$.  For baryons such as $p$ 
and $\bar{p}$, we use a string-inspired parameterization of their 
fragmentation functions as in Refs.~\cite{string-par,junction}. 
The resulting $p$ and $\bar p$ spectra are, nevertheless, sensitive to
the quark and antiquark distributions in the nucleon. 

We note that application of the energy loss calculations alone for
neutral pions at $\sqrt{s}=$62~GeV in Au+Au collisions at 
RHIC~\cite{vitev-62} yields a factor $2 - 3$ suppression with only 
a moderate dependence on $p_T$. The medium-induced 
non-Abelian bremsstrahlung thus plays an important role in 
modifying the relative importance of the pQCD  fragmentation mechanism 
when compared to the low $p_T$ non-perturbative approaches 
such as string and baryon junction mechanisms~\cite{junction,vance}, 
relativistic hydrodynamics~\cite{hydro,hydro-shuryak}, and quark 
coalescence~\cite{greco,greco2,fries,moln04}. 

\section{Particle Spectra and Ratios}\label{results}

To obtain hadron transverse momentum spectra, we evaluate the 
multidimensional integral in Eq.~(\ref{coal}) by the Monte Carlo
method using test particles that are distributed uniformly in the fireball
according to the distribution function and geometry described in 
Sects.~\ref{parameters} and~\ref{pQCD}. Details of this method can be
found in Ref.~\cite{greco2}.
As in our studies for Au+Au collision at 
center of mass energy $\sqrt{s_{NN}}=200$~GeV, 
we include quark coalescence into stable hadrons such as $\pi$, $p$,
and $\bar p$ as well as unstable resonances such as $\rho$, $\omega$, 
$\Delta$, and $K^{\star}$. Decays of the resonances are included, and they
modify the spectra of stable hadrons mainly at $ p_T <2$~GeV, leading
to a better agreement with experimental data at $\sqrt{s_{NN}}=200$~GeV.

\begin{figure}[ht]
\includegraphics[height=2.35in,width=3in,angle=0]{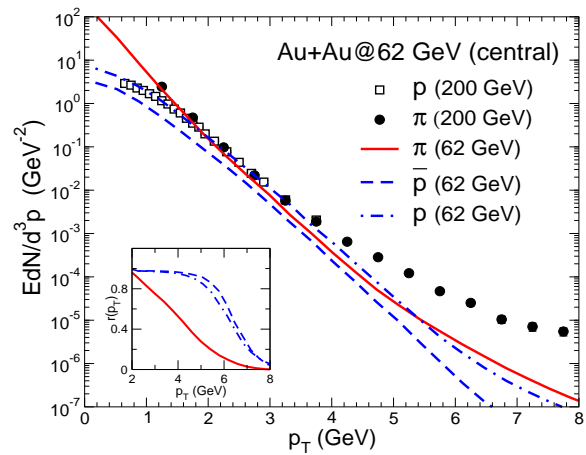}
\caption{(Color online) Hadron transverse momentum spectra from central
Au+Au collisions at $\sqrt{s_{NN}}=62$~GeV. Predicted results are
for pions (solid line), protons (dash-dotted line), and antiprotons 
(dashed line). Experimental $\pi^0$ (filled circles) and proton (open
squares) data at $\sqrt{s_{NN}}=200$~GeV \cite{phenix-spec,phenix-idp} 
are shown for reference. Inset: The ratio $r(p_T)$ of the contribution
from quark coalescence to that from both quark coalescence and minjet
fragmentation for $\pi$ (solid line), $p$ (dash-dotted line),
and $\bar p$ (dashed lixne).}
\label{spectra}
\end{figure}

In Fig.~\ref{spectra}, we show predicted $p_T$ spectra of pions
(solid line), protons (dash-dotted line) and antiprotons (dashed line)
including both coalescence and fragmentation contributions for Au+Au 
collisions at $\sqrt{s_{NN}}=62$~GeV.  For comparison, we also show 
the experimental pion (filled circles) and proton (open squares) data
at $\sqrt{s_{NN}}=200$~GeV from the PHENIX collaboration~\cite{phenix-spec}. 
At low $p_T$, both pion and proton spectra are similar at the two 
energies. This is not surprising as the total charged particle
multiplicity, which is dominated by low $p_T$ particles, is only
reduced by about 20$\%$ as the collision energy decreases from 200 
to 62~GeV. At higher $p_T$ the pion spectrum depends mostly on the 
fragmentation of quenched minijets. With respect to the case 
at 200~GeV, the quenching mechanism is somewhat less effective due to a
lower value of the initial gluon density~\cite{vitev-62}. Preliminary
data from the PHOBOS experiment on the suppression factor for charged 
particles has, indeed, confirmed such a trend~\cite{phobos-raa62}. 
On the other hand, less minijet partons are initially produced at
$\sqrt{s_{NN}}=62$~GeV than at $\sqrt{s_{NN}}=200$~GeV and their 
transverse momentum distribution is also steeper, which amplifies 
the quenching effect at high $p_T$. The interplay between the 
non-perturbative and perturbative yields results in a $\pi$ spectrum 
that is similar to the one at $\sqrt{s_{NN}}=200$~GeV around 
$p_T \simeq 2.5$~GeV but falls down more quickly as shown in 
Fig.~\ref{spectra}. 

Fig.~\ref{spectra} further shows that although pions dominate
at low $p_T$, more protons are produced at intermediate $p_T$ up to 
5.5~GeV. The antiproton yield also becomes comparable to the pion yield at 
intermediate transverse momenta. Since the minijet fragmentation
gives the proton and antiproton spectra that are about a factor of 
4 lower than the pion spectrum, the large proton and antiproton
yields at intermediate transverse momenta are mainly due to 
contributions from the quark coalescence. This is further illustrated 
in the inset of Fig.~\ref{spectra}, where the ratio $r(p_T)$ of the
contribution from quark coalescence to that from both quark 
coalescence and minijet fragmentation is shown for $\pi$ (solid line), 
$p$ (dash-dotted line), and $\bar p$ (dashed line). It is seen that the 
quark coalescence contribution to proton production at intermediate 
transverse momenta overtakes the suppressed proton spectrum from
minijet fragmentation up to $p_T < 6.5$~GeV, while for pions it 
is dominant only up to $p_T < 4$~GeV. 

\begin{figure}[ht]
\includegraphics[height=2.25in,width=3in,angle=0]{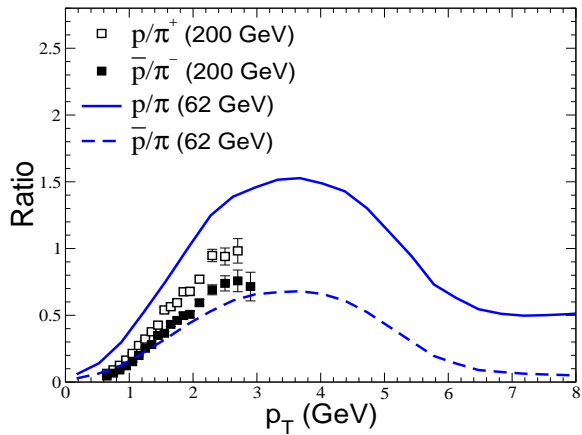}
\caption{(Color online) Same as Fig.~\ref{spectra} for the ratios of 
proton (solid line) and antiproton (dashed line) to pion spectra.}
\label{ratios}
\end{figure}

The predicted $p/\pi$ ratio in central Au+Au collisions at 
$\sqrt{s_{NN}}=62$~GeV is shown in Fig.~\ref{ratios}, and it
reaches a value of about 1.5 at $p_T \sim 4$~GeV.
For comparison, the $p/\pi$ ratio of about 1 was measured in
collisions at $\sqrt{s_{NN}}=200$~GeV as shown by open squares for 
the experimental data~\cite{phenix-idp}. The large value at lower 
collision energy is due to the larger difference between the slopes 
of the spectra from fragmentation and coalescence in collisions 
at $\sqrt{s_{NN}}=62$~GeV than that at $\sqrt{s_{NN}}=200$~GeV. 
Also shown in Fig.~\ref{ratios} is the $\bar p/\pi$ ratio at collision 
energy of $\sqrt{s_{NN}}=62$~GeV.  It can reach a maximum value of
about 0.7, which is close to the value at $\sqrt{s_{NN}}=200$~GeV, 
shown by filled squares for the experimental data, but at slightly 
larger $p_T$, as a result of the steeper $\pi$ spectrum from the 
fragmentation contribution. Because of the difference in the 
baryon chemical potentials, a larger difference between $\bar p/\pi$ 
and $p/\pi$ ratios is seen at $\sqrt{s_{NN}}=62$~GeV than at 200~GeV. 

\section{Summary}\label{summary}

Using a hadronization model  based on quark coalescence and minijet 
fragmentation, we predicted the transverse momentum spectra of pions, 
protons and antiprotons for central Au+Au collisions at 
$\sqrt {s_{NN}}=62$~GeV.  We found a larger enhancement of the $p/\pi$ 
ratio with respect to the one seen in nuclear collisions at 
$\sqrt {s_{NN}}=200$~GeV. Our result is mainly due to a steeper pion 
spectrum, that is dominated by minijet fragmentation already at 
$p_T \simeq 4$~GeV, and a proton spectrum that is instead still
dominated by contributions from the quark coalescence up 
to $p_T \simeq 6.5$~GeV. Experimental confirmation of this prediction 
will provide a stronger evidence for quark coalescence as a likely 
non-perturbative mechanism for hadronization, especially for protons 
(and in general the baryons) at transverse momenta up to about 6~GeV. 
This in turn will also facilitate the determination of the
characteristic energy at which the quark-gluon plasma is formed in 
heavy-ion reactions.

Our results depend on the properties of the bulk partonic matter at
the moment of hadronization and the amount of minijet quenching. 
In the present study, they are determined from extrapolations based 
on  available data at $\sqrt{s_{NN}}=17$~GeV from SPS as well as 
$\sqrt{s_{NN}}=130$ and $200$~GeV from RHIC. With upcoming
experimental data on the total transverse energy, baryon chemical 
potential, and  minijet  suppression factor in collisions at 
$\sqrt{s_{NN}}=62$~GeV, more reliable predictions on hadron spectra 
and their ratios can be obtained. Our work, however, represents
an important first step in investigating the interplay between  
the non-perturbative and perturbative hadron production mechanisms 
at intermediate RHIC energies.  

\vspace{0.5cm}
\begin{acknowledgments}

We are grateful to Zhangbu Xu for helpful comments on an earlier
version of this paper. This paper was based on work supported by the 
US National Science Foundation under Grant No. PHY-0098805 and the 
Welch Foundation under Grant No. A-1358 (C.M.K. and V.G.) and the J.R.
Oppenheimer fellowship of the Los Alamos National Laboratory (I.V.)

\end{acknowledgments}

\end{document}